\documentclass[showpacs,twocolumn,aps,superscriptaddress]{revtex4}
\usepackage{amsfonts}
\usepackage{amsmath}
\usepackage{amssymb}
\usepackage{graphicx}
\usepackage{verbatim}
\usepackage{textcomp}

\providecommand{\U}[1]{\protect\rule{.1in}{.1in}}

\begin{document}
\title{On-chip scalable optomechanical magnetometers}

\author{Bei-Bei Li}
\affiliation{Center for Engineered Quantum Systems, School of Mathematics and Physics, The University of Queensland, St Lucia, Queensland 4072, Australia.}
\author{Douglas Bulla}
\affiliation{Defense Science and Technology Group, Department of Defense, Edinburgh, SA  5111, Australia.}
\author{Varun Prakash}
\affiliation{Center for Engineered Quantum Systems, School of Mathematics and Physics, The University of Queensland, St Lucia, Queensland 4072, Australia.}
\author{Stefan Forstner}
\affiliation{Center for Engineered Quantum Systems, School of Mathematics and Physics, The University of Queensland, St Lucia, Queensland 4072, Australia.}
\author{Ali Dehghan-Manshadi}
\affiliation{School of Mechanical and Mining Engineering, The University of Queensland, St Lucia, Queensland 4072, Australia.}
\author{Halina Rubinsztein-Dunlop}
\affiliation{Center for Engineered Quantum Systems, School of Mathematics and Physics, The University of Queensland, St Lucia, Queensland 4072, Australia.}
\author{Scott Foster}
\affiliation{Defense Science and Technology Group, Department of Defense, Edinburgh, SA  5111, Australia.}
\author{Warwick P. Bowen}
\email{w.bowen@uq.edu.au}
\affiliation{Center for Engineered Quantum Systems, School of Mathematics and Physics, The University of Queensland, St Lucia, Queensland 4072, Australia.}

\date{\today}

\begin{abstract}

The dual-resonant enhancement of mechanical and optical response in cavity optomechanical magnetometers enables precision sensing of magnetic fields. In previous working prototypes of such magnetometers, a cavity optomechanical system is functionalized by manually epoxy-bonding a grain of magnetostrictive material. While this approach allows proof-of-principle demonstrations, practical applications require more scalable and reproducible fabrication pathways. In this work, we scalably fabricate optomechanical magnetometers on a silicon chip, with reproducible performance across different devices, by sputter coating a magnetostrictive film onto high quality toroidal microresonators. Furthermore, we demonstrate that thermally annealing the sputtered film can improve the magnetometer sensitivity by a factor of 6.3. A peak sensitivity of 585 pT/$\rm{\sqrt{Hz}}$ is achieved, which is comparable with previously reported results using epoxy-bonding.

\end{abstract}



\maketitle

\section{Introduction}




Ultrasensitive magnetometers are key components for various applications, such as magnetic anomaly detection \cite{MAD1,MAD2}, mineral exploration \cite{mine1,mine2}, magnetoencephalography \cite{MEG1,MEG2}, and magnetic resonance imaging \cite{NMR1,NMR2}. Currently, the most advanced commercial magnetometer the superconducting quantum interference device (SQUID) based magnetometer \cite{SQUID1,SQUID2,SQUID3}. However, the requirement of cryogenic cooling increases the complexity of SQUID magnetometer systems. To circumvent this requirement, various high precision magnetometers operating at room temperature have been developed \cite{atomic1,atomic2,NV1,NV2,cavity1,cavity2,cavity3,cavity4}. Among them, the cavity optomechanical magnetometer \cite{cavity1,cavity2,cavity3,cavity4} offers the advantages of small size, weight, and power consumption; ease of on-chip integration; high sensitivity; and broad bandwidth. In such magnetometers, the strain induced by a magnetic field on an embedded magnetostrictive material deforms the optical cavity. The resulting cavity resonance frequency shift is read out optically. The combination of resonance-enhanced mechanical and optical response \cite{2008Science,2014RMP,Warwick book} enables unprecedented transduction sensitivity \cite{attoNewton,attometer}, surpassing that of the previously demonstrated magnetostrictive magnetometers by several orders of magnitude \cite{magnetostrictive1,magnetostrictive2,magnetostrictive3,magnetostrictive4}.


The first working prototype of a cavity optomechanical magnetometer was realized by manually depositing a grain of the magnetostrictive material Terfenol-D on top of a microcavity, and affixing it using epoxy. This magnetometer achieved a sensitivity of hundreds of nT/$\sqrt{\mathrm{Hz}}$ \cite{cavity1}. The sensitivity was further improved, to a level of hundreds of pT/$\sqrt{\mathrm{Hz}}$, by fabricating a central hole in the cavity structure and depositing Terfenol-D into it \cite{cavity2}. However, the manual deposition process requires precise positioning of micro-sized Terfenol-D grains relative to the microcavity. Combined with the use of epoxy-bonding, this makes the approach ill-suited for scalable fabrication. Furthermore, reproducible performance across devices is hard to realize, due to the random geometry of the Terfenol-D grain in each device. To overcome these challenges, in this work we develop a controllable fabrication method, by deterministically sputter coating thin films of Terfenol-D into the microcavities. With this method, the Terfenol-D deposition for all the devices in a complete wafer can be performed in one run. The sensitivity of 10 devices is characterized, showing quite similar performances across devices. A peak sensitivity of 585~pT/$\rm{\sqrt{Hz}}$ is achieved, which is comparable with previously reported devices fabricated using the manual deposition method \cite{cavity2}. Furthermore, by thermally annealing the sputtered Terfenol-D film, we show that the magnetostrictive coefficient can be increased by a factor of 6.3, leading to improvement in the sensitivity. This sputter coating method provides a scalable and reproducible fabrication pathway for cavity optomechanical magnetometers on a silicon chip.


\section{Device design and fabrication}

\begin{figure}[t!]
\begin{center}
\includegraphics[width=8.5cm]{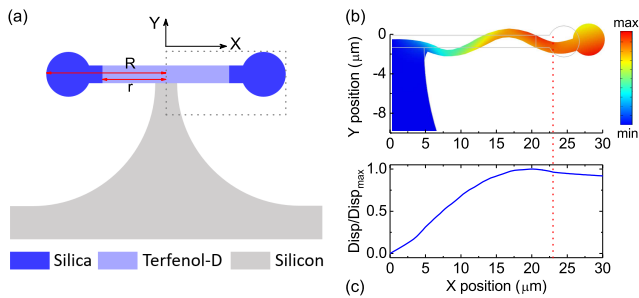}
\end{center}
\caption{(Color online) (a) Schematic of the side view of the designed magnetometer, which consists of a silica microtoroid with Terfenol-D embedded inside, sitting on a silicon pedestal. (b) the mode profile of the radial breathing mode, within the dotted region in (a). The color map shows the displacement of the disk along the radial direction. The grey border shows the equilibrium position of the disk. (c) the displacement distribution of the radial breathing mode (shown in (b)), normalized to the maximum value. The red dotted line denotes the position of the boundary between the Terfenol-D and silica.}
\label{fig1}
\end{figure}

Figure \ref{fig1}(a) shows a schematic side view of the designed magnetometer. It consists of a silica microtoroid \cite{toroid} with a Terfenol-D disk embedded inside, supported by a silicon pedestal. The silica toroid supports high quality (\emph{Q}) whispering gallery modes that confine optical field around its circumference. The system also supports high \emph{Q} mechanical resonance modes, e.g., radial breathing mode (Fig. \ref{fig1}(b)). The mechanical motion modifies the circumference of the optical cavity, and thus shifts the optical resonance. In this magnetometer system, the Terfenol-D deforms in the presence of a magnetic field and exerts a force on the silica resonator at the boundary between Terfenol-D and silica. To maximize the force we make the Terfenol-D disk as large as possible given fabrication constrains. For a magnetometer with toroid outer radius of $R$=30~\textmu m and Tefenol-D radius of $r$=23~\textmu m, the cross sectional profile of the radial breathing mode (the region within the dotted frame shown in \ref{fig1}(a)) is shown in Fig. \ref{fig1}(b), obtained using COMSOL Multiphysics. For this mode, the displacement distribution of the magnetometer along radial direction is plotted in Fig. \ref{fig1}(c). From both Figs. \ref{fig1}(b) and (c), it can be seen that the displacement at the boundary between Terfenol-D and silica where the magnetic field induced force is exerted reaches a maximum. As a result, the deformation of Terfenol-D can effectively drive the mechanical motion of the silica resonator. In the simulation, the thickness of the Terfenol-D film is set to be 2~\textmu m, and the radius of the silicon pedestal is 5~\textmu m.


\begin{figure}[t!]
\begin{center}
\includegraphics[width=7.5cm]{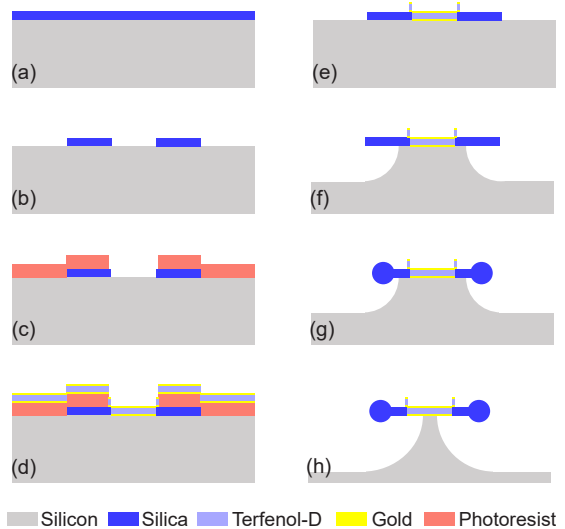}
\end{center}
\caption{(Color online) Schematic of the cavity optomechanical magnetometer fabrication process.}
\label{fig2}
\end{figure}

The magnetometer fabrication process is shown in Fig. \ref{fig2}. Starting from a silicon wafer (thickness of 500~\textmu m) with a 2~\textmu m-thick thermally oxidized silica layer on top (shown in (a)), silica annular disks are patterned with standard photolithography using a positive photoresist AZ 1518, followed by a hydrofluoric (HF) acid etching process (step (b)). Next a Terfenol-D thin film with similar thickness as the silica is sputter coated into the hole in each annular disk. For this, first the negative photoresist AZ 2070 is spin coated over the wafer, with the photoresist over the holes removed using a second photolithography step (step (c)). In this step, the radius of the opening in the photoresist is designed to be 5~\textmu m larger than the holes on each side, to leave some tolerance for alignment in the photolithography process. This also leaves some overlap between Terfenol-D and silica to increase the bonding between them. Then a layer of 2~\textmu m-thick Terfenol-D film is sputter coated on top of the wafer (step (d)). Note that a 10~nm-thick gold layer is coated both below and above the Terfenol-D to protect it from oxidization; oxidization can spoil the magnetostriction and degrade the sensitivity of the magnetometer \cite{handbook}. The excess Terfenol-D and gold outside the holes are removed through a lift-off process using acetone (step (e)). A xenon difluoride (XeF$_{2}$) etch is then performed to undercut the silicon pedestal (step (f)), followed by a carbon dioxide (CO$_2$) laser reflow process to melt the edge of the silica disk to form a toroid structure and increase the optical \emph{Q} factor of the silica microcavity (step (g)). Finally a second XeF$_2$ etch is performed to further underetch the silicon pedestal (step (h)), in order to release the Terfenol-D from the silicon pedestal for better magnetic actuation. Note that in step (f), the silicon pedestal is kept $\sim5$~\textmu m larger than the Terfenol-D disk on each side, to increase the thermal conductivity out of the Terfenol-D layer during the CO$_2$ laser reflow process, and thereby minimize the thermal load on the Terfenol-D.

\section{Device characterization and measurement}

\begin{figure}[t!]
\begin{center}
\includegraphics[width=7.5cm]{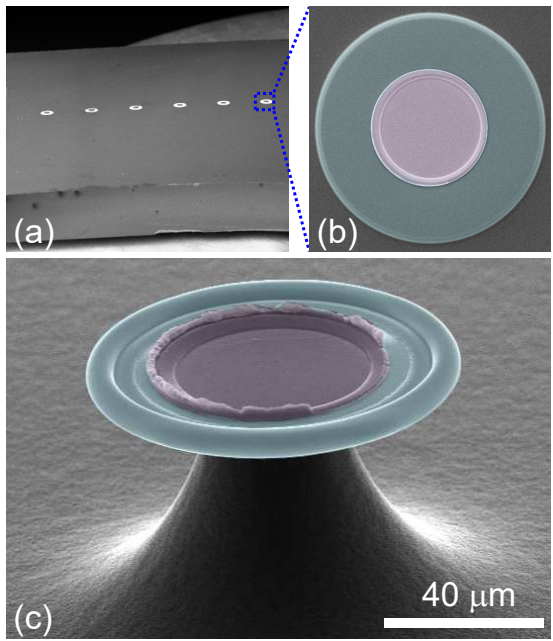}
\end{center}
\caption{(Color online) Scanning electron microscope (SEM) images of the fabricated devices. (a) a silicon chip with an array of silica disks embedded with Terfenol-D. (b) Zoom-in of one of the silica disks (prior to the first XeF$_2$ etching process) with Terfenol-D embedded in the center. (c) The final device: a toroid microresonator with a Terfenol-D disk embedded in the center.}
\label{fig3}
\end{figure}

A scanning electron microscope (SEM) image of a silicon chip with an array of silica microdisks with Terfenol-D disks embedded inside is shown in Fig. \ref{fig3}(a). A magnified image of one disk is shown in (b), exhibiting good uniformity of the Terfenol-D film. The thickness of the Terfenol-D film is measured to be $\sim2.2$~\textmu m. The final magnetometer device is shown in Fig. \ref{fig3}(c), with toroid diameter of $\sim$80~\textmu m.

To investigate the optical properties of the fabricated devices, we couple light from a laser in the 1550~nm wavelength band into the microtoroid through a tapered fiber with diameter of $\sim$1~\textmu m. The transmitted light is detected using a 125~MHz-bandwidth InGaAs photoreceiver. The wavelength of the light from the laser is swept to find the optical resonance modes. Typical optical \emph{Q} factors are found to be in the range of $Q$ factors of $10^7$ to $10^8$, similar to that of pure silica toroids \cite{toroid}, indicating that the presence of Terfenol-D has no significant influence on the optical attenuation.

\begin{figure}[t!]
\begin{center}
\includegraphics[width=7.5cm]{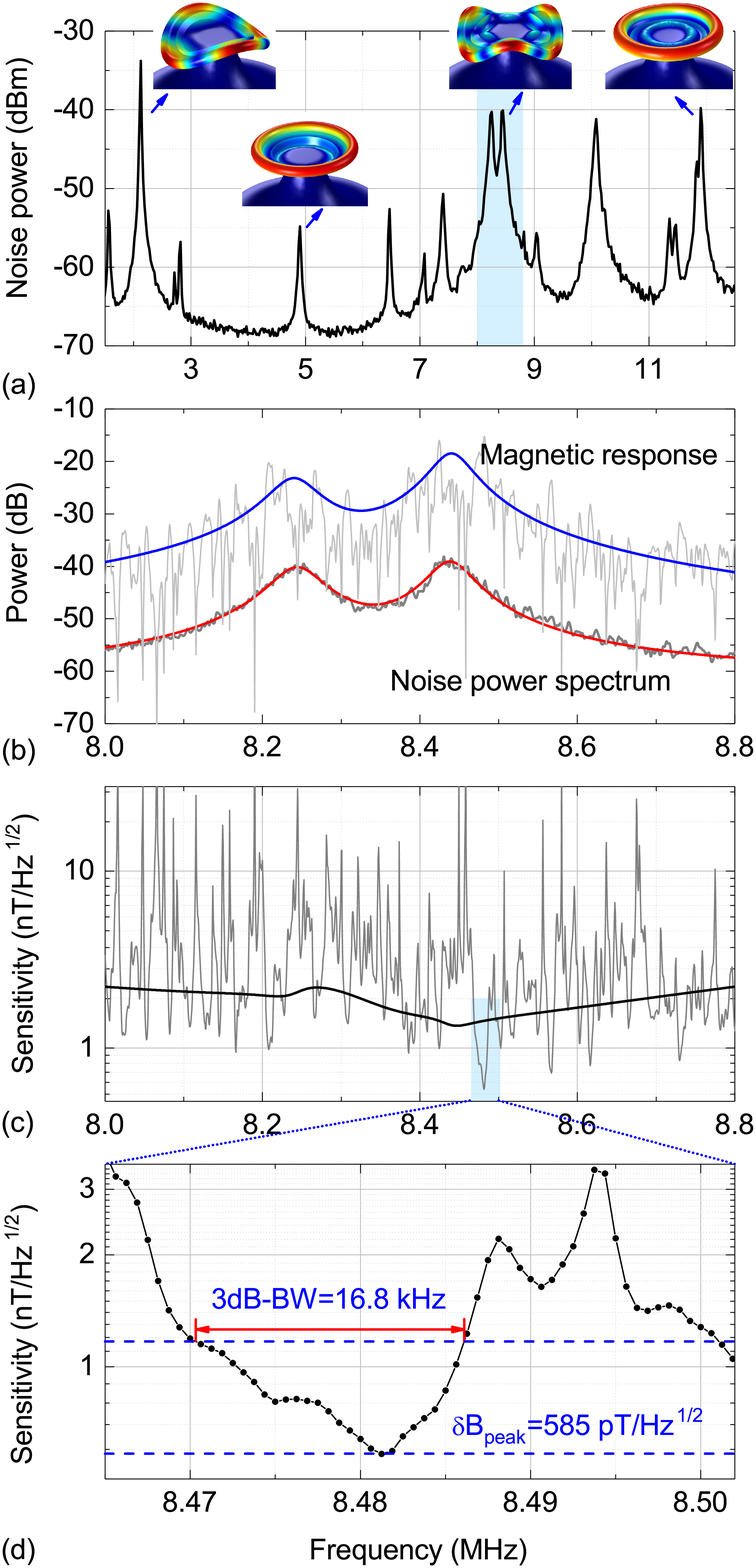}
\end{center}
\caption{(Color online) (a) Noise power spectrum of a magnetometer, with the peaks representing the thermally excited mechanical modes. The profiles of four of the modes are shown in the inset. (b) Zoom-in of the noise power spectrum (dark grey curve) in the frequency range 8~MHz-8.8~ MHz (the blue shadowed region in (a)), and the magnetic response (light grey curve) in the same frequency range. The red curve is the fitted result for the measured noise power spectrum. The blue curve is the envelop of the magnetic response spectrum, following the mechanical resonance spectrum. (c) Sensitivity spectrum in the frequency range of 8~MHz-8.8~MHz. The black curve is the predicted sensitivity in the absence of the sharp magnetostrictive resonances, derived from the fitted data in (b). (d) Zoom-in on the sensitivity spectrum in the frequency range of 8.465~MHz-8.502~MHz, with the peak sensitivity of $\sim585$ pT/$\sqrt{\rm{Hz}}$ and the 3-dB bandwidth of 16.8~kHz.}
\label{fig4}
\end{figure}

The mechanical motion of the microresonator translates into a periodic modulation of the intracavity field. Intensity (phase) modulation can be optically read out by locking the frequency of the probe light on the side (center) of the cavity resonance mode. In our experiment, the detuning of the laser is stabilized by thermally locking the laser at half maximum of the transmission on the blue-detuned side of an optical resonance \cite{locking}. The output probe light, captured by the tapered fiber, is detected by a photoreceiver. The direct current (DC) part of the photocurrent from the photoreceiver is sent to an oscilloscope to monitor the transmission. The alternating current (AC) part is sent to an electronic spectrum analyzer (ESA) to measure the mechanical spectrum of the cavity. To test the magnetic response of the magnetometer, a solenoid is used to produce a magnetic field with a known strength. The frequency of the produced magnetic field can be swept by sweeping the frequency of the voltage applied to the coil from the output port of an electronic network analyzer (ENA), and the magnetic response of the magnetometer at each frequency is measured with the input port of the same ENA.

In such a cavity optomechanical magnetometer system, the main noise sources consist of the thermal noise $N_{\mathrm{th}}$ from the environment and the shot noise $N_s$ from the probe laser. The thermal noise is proportional to the mechanical susceptibility, $\chi=1/[m(\omega^2-\Omega^2-i \Gamma)]$ for a single mechanical mode case, with $m$ being the effective mass of the mechanical oscillator, and $\Omega$ and $\Gamma$ its resonance frequency and damping rate. The thermal noise reaches a maximum on mechanical resonance and decays rapidly when detuned off resonance, whereas the shot noise does not depend on the frequency. As a result, the total noise is typically dominated by the thermal noise on mechanical resonance, and by shot noise in the off-resonance frequency ranges. The magnetic field induced signal $S$ also follows $\chi$. As both $N_{\mathrm{th}}$ and $S$ are proportional to the probe power, and $N_{s}$ is proportional to the squre-root of power, increasing the probe power reduces the relative contribution from shot noise, increasing the signal to noise ratio $\mathrm{SNR}=S/(N_{\mathrm{th}}+N_{s})$, and therefore provides a better magnetic field sensitivity. At sufficiently high probe power the sensitivity will saturate at the thermal-noise-limited sensitivity. In our experiment, the thermal noise dominant regime is reached for most of the measured frequency range (0.5-60~MHz), with a probe power of only $\sim$50~\textmu W. Figure \ref{fig4}(a) shows the noise power spectrum of one of the fabricated magnetometers in the frequency range from 1.5~MHz-12.5~MHz. Multiple peaks can be seen in the spectrum, which correspond to thermally excited mechanical modes. The profiles of four of the mechanical modes are shown in the inset, obtained using COMSOL Multiphysics. A zoom-in of the noise power spectrum in the frequency range of 8~MHz-8.8~MHz is shown in the dark grey curve in Fig. \ref{fig4}(b), fitting well to a pair of Lorentzian peaks (red curve).

The magnetic field sensitivity of the magnetometer is then characterized. A magnetic field with known strength $B_{\mathrm{ref}}$ at a certain frequency $\omega_{\mathrm{ref}}$ is first applied to the magnetometer, to obtain the sensitivity at this frequency $\delta B(\omega_\mathrm{ref})$. In order to obtain the magnetic response across the full frequency range, the frequency of the magnetic field is swept using the ENA. The magnetic response is found to vary significantly over frequency ranges of $\sim$10~kHz, as shown in the light grey curve in Fig. \ref{fig4}(b), consistent with previously reported results, and attributed to sharp magnetostrictive resonances \cite{cavity1}. Modulo these steep resonances, the envelope of the magnetic response as a function of frequency follows the mechanical resonance spectrum, as shown by the blue curve. With the noise power spectrum $N(\omega)$ and the magnetic response spectrum $R(\omega)$, the sensitivity $\delta B(\omega)$ of the magnetometer in this frequency range can be derived from $\delta B(\omega)=\delta B(\omega_{\mathrm{ref}})\sqrt{[N(\omega)R(\omega_{\mathrm{ref}})]/[N(\omega_{\mathrm{ref}})R(\omega)]}$ \cite{cavity1}, as shown in the grey curve in Fig. \ref{fig4}(c). The black curve in Fig. \ref{fig4}(c) is the predicted sensitivity that would be obtained in the absence of magnetostrictive resonances, derived using the data in the red and blue curves in Fig. \ref{fig4}(b). A zoom-in of the sensitivity spectrum in the frequency range of 8.465~MHz-8.502~MHz is shown in Fig. \ref{fig4}(d), with the peak sensitivity of $\sim585$ pT/$\sqrt{\rm{Hz}}$ and the 3-dB bandwidth of 16.8~kHz.


Studies show that the magnetostrictive coefficient of Terfenol-D is very sensitive to its composition \cite{handbook}. In our experiment, the atomic composition of the sputter coated Terfenol-D film is measured via energy dispersive X-ray spectroscopy (EDS), to be 74\% Fe, 8\% Tb, and 18\% Dy, which deviates from the ideal composition: 66\% Fe, 10\% Tb, and 24\% Dy. Further optimization of the composition through optimizing the sputter coating parameters is desirable to further improve the sensitivity.

\begin{figure}[t!]
\begin{center}
\includegraphics[width=7.5cm]{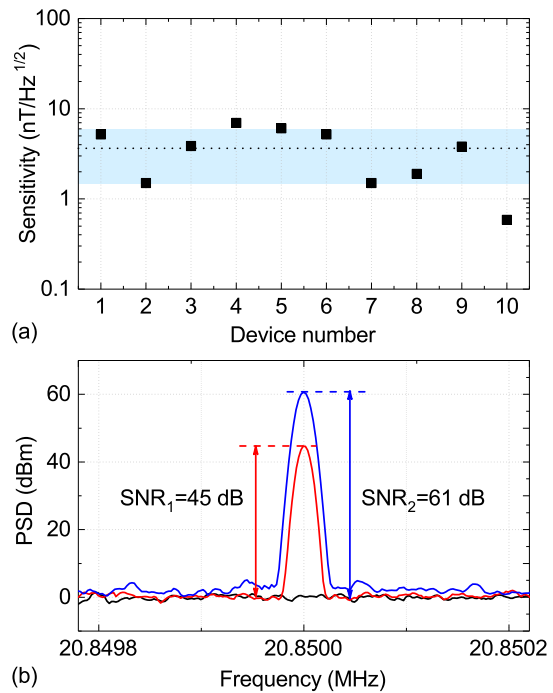}
\end{center}
\caption{(Color online) (a) Peak sensitivities of 10 different magnetometers, showing reproducible sensitivities. The dotted line and the shaded area represent the mean and standard deviation. (b) The noise power spectra for a magnetometer, driven with a magnetic field at a frequency of 28.5~MHz, showing an improvement of SNR by $\sim$16~dB through annealing process. The red and blue curves are the data before and after annealing, respectively.}
\label{fig5}
\end{figure}

Due to the parallel and reproducible fabrication method, it is found that all magnetometers fabricated in this way show quite similar sensitivities. In Fig. \ref{fig5}(a) we plot the peak sensitivity of 10 different magnetometers fabricated on the same wafer, with sensitivity values ranging from 0.585~nT/$\sqrt{\rm{Hz}}$ to 7 nT/$\sqrt{\rm{Hz}}$. The mean sensitivity achieved is 3.65~nT/$\sqrt{\mathrm{Hz}}$ (the dotted line) with a standard deviation of 2.19~nT/$\sqrt{\mathrm{Hz}}$ (the shaded area). As a comparison, in our experience the sensitivities of magnetometers fabricated with manual Terfenol-D deposition method in previous works can vary by more than two order of magnitude.

It has been shown that sputter coated Terfenol-D is amorphous which results in an order of magnitude smaller magnetostrictive coefficient than crystalline Terfenol-D \cite{sputter}. Two methods could be used to make the Terfenol-D film crystalline: increasing the temperature of the substrate to 400~$^\circ$C during the sputter coating process, or post annealing the amorphous Terfenol-D film at 400~$^\circ$C \cite{sputter}. In our case, the first method is not applicable, as the photoresist on the wafer cannot survive 400~$^\circ$C, whereas the post annealing is suitable. We anneal the fabricated magnetometers at 400~$^\circ$C for 6 hours at high vacuum (about 10$^{-7}$ Torr). Besides crystallizing the sputtered film, the thermal annealing also releases the stress in the film, which may affect the magnetostriction \cite{sputter}. As shown in the power spectra of a magnetometer before and after the thermal annealing, driven by a magnetic field with a known strength at a frequency of 28.5~MHz, the SNR is improved by 16~dB after annealing, corresponding to an enhancement in sensitivity by a factor of 6.3.

\section{Conclusion}

In summary, we have scalably fabricated on-chip cavity optomechanical magnetometers with reproducible sensitivities across devices, by sputter coating Terfenol-D thin films into high-$Q$ microresonators. A peak sensitivity of 585~pT/$\sqrt{\mathrm{Hz}}$ is achieved, comparable to the previously reported result using manual Terfenol-D deposition method. It is also demonstrated that annealing the sputter coated Terfenol-D film can improve the magnetostrictive coefficient and therefore the sensitivity of the magnetometer, by a factor of 6.3. Further optimization of the sputter coating parameters is desirable to further improve the sensitivities. This sputter coating method provides a scalable and reproducible fabrication pathway for the chip integrated magnetometers, opening up possibilities for applications, such as on-chip microfluidic nuclear magnetic resonance and magnetoencephalography.


\begin{acknowledgments}
We thank James Bennett for the helpful discussions on the manuscript. We thank support from the DARPA QuARSAR Program, Australian Research Council projects (DP140100734 and FT140100650), and Australian Defence Science and Technology
Group projects (CERA49 and CERA50). Bei-Bei Li also acknowledges the support from the University of Queensland Postdoctoral Research Fellowship (No. 2014001447). With the exception of sputter coating and laser reflow, device fabrication was performed within the Queensland Node of the Australian Nanofabrication Facility.

\end{acknowledgments}


\begin{thebibliography}{99}

\bibitem{MAD1} C. Li, S. Huang, D. Wei, Y. Zhong, and K. Y. Gong, Journal of Engineering Science and Technology Review \textbf{8}, 105 (2015).

\bibitem{MAD2} A. Sheinker, L. Frumkis, B. Ginzburg, N. Salomonski, and B.-Z. Kaplan, IEEE Transactions on Magnetics \textbf{45}, 160 (2009).

\bibitem{mine1} A. Edelstein, J. Phys. Condens. Matter \textbf{19}, 165217 (2007).

\bibitem{mine2} H. G. Meyer, R. Stolz, A. Chwala, and M. Schulz,  Phys. Status Solidi (a) \textbf{2}, 1504 (2005).

\bibitem{MEG1} Elena Boto, Niall holmes, James Leggett, Gillian roberts, Vishal Shah, Sofie S. meyer, Leonardo Duque mu\~{a}oz, Karen J. mullinger, Tim m. Tierney, Sven Bestmann, Gareth r. Barnes, Richard Bowtell, and matthew J. Brookes, Nature \textbf{555}, 657 (2018).

\bibitem{MEG2} H. Xia, A. Ben-Amar Baranga, D. Hoffman, and M. V. Romalis. Appl. Phys. Lett. \textbf{89}, 211104 (2006).

\bibitem{NMR1} David r. Glenn, Dominik B. Bucher, Junghyun Lee, mikhail D. Lukin, hongkun Park, and Ronald L. Walsworth, Nature \textbf{555}, 351 (2018).

\bibitem{NMR2} I. Savukov and T. Karaulanov, Appl. Phys. Lett. \textbf{103}, 043703 (2013).

\bibitem{SQUID1} J. R. Kirtley, M. B. Ketchen, K. G. Stawiasz, J. Z. Sun, W. J. Gallagher, S. H. Blanton, S. J. Wind, Appl. Phys. Lett. \textbf{66}, 9 (1995).

\bibitem{SQUID2} F. Baudenbacher, L. E. Fong, J. R. Holzer, M. Radparvar, Appl. Phys. Lett. \textbf{82}, 20 (2003).

\bibitem{SQUID3} M I Faley, U Poppe, K Urban, D N Paulson, and R L Fagaly, Journal of Physics: Conference Series \textbf{43}, 11991202 (2006).

\bibitem{atomic1} H. B. Dang, A. C. Maloof, and M. V. Romalis, Appl. Phys. Lett. \textbf{97}, 151110 (2010).

\bibitem{atomic2} M. Vengalattore, J. M. Higbie, S. R. Leslie, J. Guzman, L. E. Sadler, and D. M. Stamper-Kurn, Phys. Rev. Lett. \textbf{98}, 200801 (2007).

\bibitem{NV1} G. Balasubramanian, P. Neumann, D. Twitchen, M. Markham, R. Kolesov, N. Mizuochi, J. Isoya, J. Achard, J. Beck, J. Tissler, V. Jacques, P. R. Hemmer, F. Jelezko, and J. Wrachtrup. Nat. Matter. \textbf{8}, 383 (2009).

 \bibitem{NV2} T. Wolf, P. Neumann, K. Nakamura, H. Sumiya, T. Ohshima, J. Isoya, and J. Wrachtrup. Phys. Rev. X \textbf{5}, 041001 (2015).

\bibitem{cavity1} S. Forstner, S. Prams, J. Knittel, E. D. vanOoijen, J. D. Swaim, G. I. Harris, A. Szorkovszky, W. P. Bowen, H. Rubinsztein-Dunlop, Phys. Rev. Lett. \textbf{108}, 120801 (2012).

\bibitem{cavity2} S. Forstner, E. Sheridan, J. Knittel, C. L. Humphreys, G. A. Brawley, H. Rubinsztein-Dunlop, W. P. Bowen, Adv. Mater. \textbf{26}, 6348 (2014).

\bibitem{cavity3} C. Yu, J. Janousek, E. Sheridan, D. L. McAuslan, H. Rubinsztein-Dunlop, P. K. Lam, Y. Zhang, and W. P. Bowen, Phys. Rev. Appl. \textbf{5}, 044007 (2016).


\bibitem{cavity4} J. Zhu, G. Zhao, I. Savukov, and L. Yang, Sci. Rep. \textbf{7}, 8896 (2017).

\bibitem{2008Science} T. J. Kippenberg, and K. J. Vahala, \textit{Cavity Optomechanics: Back-Action at the Mesoscale}, Science \textbf{321}, 1172 (2008).

\bibitem{2014RMP} M. Aspelmeyer, T. J. Kippenberg, and F. Marquardt, \textit{Cavity optomechanics}, Rev. Mod. Phys. \textbf{86}, 1391 (2014).

\bibitem{Warwick book} W. P. Bowen, and G. Milburn, \textit{Quantum Optomechanics} (CRC Press, Boca Raton, 2016).


\bibitem{attoNewton} E. Gavartin , P. Verlot , T. J. Kippenberg , Nat. Nanotechnol. \textbf{7}, 509 (2012).

\bibitem{attometer} K. H. Lee , T. G. McRae , G. I. Harris , J. Knittel , W. P. Bowen , Phys. Rev. Lett. \textbf{104}, 123604 (2010).

\bibitem{magnetostrictive1} F. Bucholtz, D. M. Dagenais, and K. P. Koo, Electron. Lett. \textbf{25}, 1719 (1989).

\bibitem{magnetostrictive2} R. Osiander, S. A. Ecelberger, R. B. Givens, D. K. Wickenden, J. C. Murphy, and T. J. Kistenmacher, Appl. Phys. Lett. \textbf{69}, 2930 (1996).

\bibitem{magnetostrictive3} S. Dong, J.-F. Li, and D. Viehland, Appl. Phys. Lett. \textbf{83}, 2265 (2003).

\bibitem{magnetostrictive4} Y. Hui, T. Nan, Nian, X. Sun, and M. Rinaldi, J. Microelectromech. Syst. \textbf{24}, 134 (2014).

\bibitem{toroid} D. K. Armani, T. J. Kippenberg, S. M. Spillane, and K. J. Vahala, Nature \textbf{42}, 925 (2003).

\bibitem{handbook} G. Engdahl, \textit{Handbook of Giant Magnetostrictive Materials} (Academic Press, 2000).

\bibitem{locking} T. G. McRae, Kwam H. Lee, M. McGovern, D. Gwyther, and W. P. Bowen, Opt. Express \textbf{17}, 21977 (2009).

\bibitem{sputter} K. P. Mohanchandra, S. V. Prikhodko, K. P. Wetzlar, W. Y. Sun, P. Nordeen, and G. P. Carman, AIP Adv. \textbf{5}, 097119 (2015)















\end{thebibliography}
\end{document}